\documentclass[
showpacs,
aps,prl,amsmath,amssymb,
twocolumn,
groupaddress
]{revtex4}

\usepackage{psfrag}

\def\XXint#1#2#3{{\setbox0=\hbox{$#1{#2#3}{\int}$}
     \vcenter{\hbox{$#2#3$}}\kern-.5\wd0}}

\usepackage{graphicx}

\newcommand{\be}{\begin{equation}}
\newcommand{\ee}{\end{equation}}
\newcommand{\bea}{\begin{eqnarray}}
\newcommand{\eea}{\end{eqnarray}}
\newcommand{\bs}{\begin{split}}
\newcommand{\bes}{\begin{equation}\begin{split}}
\newcommand{\ees}{\end{split} \end{equation}}
\newcommand{\es}{\end{split}}

\begin{document}

\title{Single particle spectrum of resonant population imbalanced Fermi gases
}

\author{F. Fumarola$^1$ and E. J. Mueller$^2$}
\affiliation{$^1$Physics Department, Columbia University, New York,
NY 10027, USA. \\ $^2$LASSP, Cornell University, Ithaca, NY 14853,
USA. }

\pacs{03.75.Ss
,05.30.Fk
}

\begin{abstract}
We use a T-matrix approximation to calculate the single particle
spectrum of the normal state of a gas of Fermionic atoms at low
temperature.
 In the strongly interacting regime of the polarized gas, we find that the spectrum is separated in two branches, leading to a double-peaked radiofrequency spectral feature.
\end{abstract}
\date{\today}
\maketitle

The experimental observation of fermionic superfluidity in ultracold alkali
gases \cite{superfluid} is
 one of the most important achievements in atomic
physics in recent years. By applying a magnetic field  \cite{FF1, FF2},
experimentalists have been able
to control the interaction strength and observed how superfluidity evolves
 as one increases the attraction between particles from the weak-coupling
 BCS limit.  As was predicted by theory, they saw that the superfluid state
 of weakly bound pairs continuously evolves to one described as a Bose
 Einstein condensate (BEC) of tightly bound molecules \cite{review}.
In the intermediate region the scattering amplitude saturates the bounds
 set by the unitarity of the S-matrix, and a strongly interacting superfluid
  with universal properties
appears.
The mechanism at the root of
fermionic superfluidity is the pairing between atoms in two different states,
which we will refer to as $\uparrow$ and $\downarrow$ spin. In recent
experiments \cite{ketterle,hulet}, where the spin relaxation time exceeds
 all other experimental timescales, researchers have created spin imbalanced
  Fermi gases, where the ratio of the number of atoms in each spin state
   $N_{\downarrow}/N_{\uparrow}=x$ is below one.  This population imbalance
   suppresses superfluidity \cite{Cl,Ch}, allowing one to study a low
   temperature normal phase.  Unlike the superfluid phase, where symmetries
   determine most of the systems behavior, the normal phase could be quite exotic.  Here we use a T-matrix approximation to
study the single particle spectrum of the normal state of an
imbalanced Fermi gas, calculating thermodynamic properties and
experimental observables.

We will be particularly focused on radio frequency (RF)
spectroscopy, where one uses radio waves to drive a transition from
one of the two spin states (say $\downarrow$) to a third, excited
state $|{\rm ex}\rangle$. By exploring the rate of transfer as a
function of the frequency of the radio waves, one can gain
information about the many-body density matrix. This technique was
used to measure the binding energy of pairs in the BEC regime
\cite{pairs}, and extended across the resonance to learn about the
pairing gap of unitary gas \cite{chin}. Recent measurements on
imbalanced gases of $^6$Li atoms have shown the remarkable feature
that the RF spectroscopy is qualitatively unchanged by polarizing
the gas \cite{Schunck}.  At high temperature (in the normal state of
either the equal spin gas or the spin imbalanced gas) one sees a
single peak in the RF spectrum.  As the temperature is lowered a
second peak appears.  In both the unpolarized and spin imbalanced
cases the appearance of this second peak occurs at a temperature of
order the superfluid transition temperature of the gas with equal
spin populations.  At lower temperatures yet, the original peak
disappears.  These results are surprising.  Although the unpolarized
gas undergoes a superfluid phase transition, the spin imbalanced gas
remains in the normal state. How can the normal state of the
imbalanced gas have an RF spectrum which is qualitatively
indistinguishable from that of the superfluid state of the equal
spin gas?  Is the ground state of the spin imbalanced gas an exotic
state containing noncondensed bosonic pairs?

\begin{figure}[!tb]
\includegraphics[height=2in,width=3.2in,angle=0]{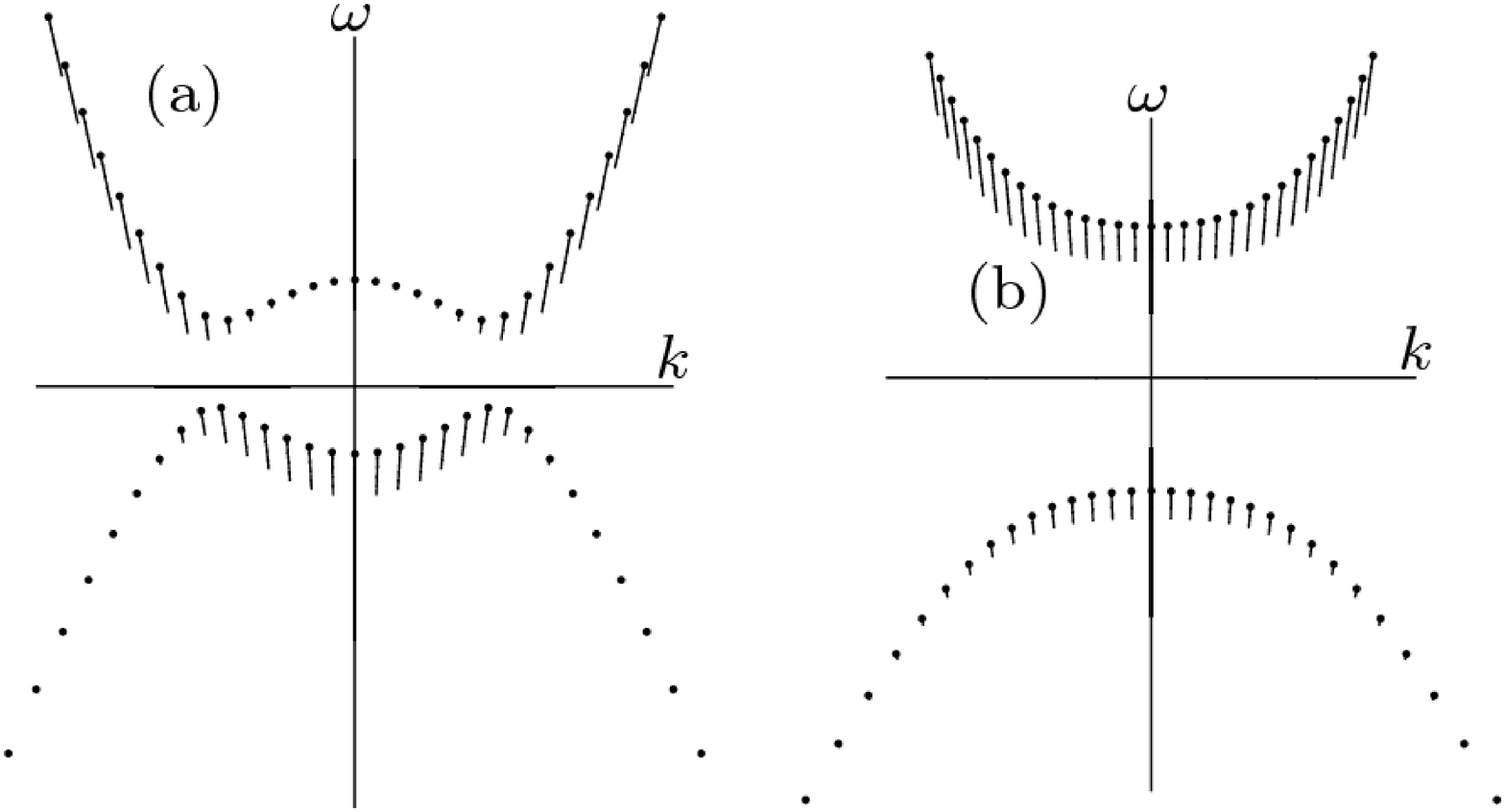}
\caption{Three dimensional depiction of mean field spectral density
of an unpolarized superfluid gas: (a) ``BCS regime",
$\Delta=0.5\mu$, (b) ``BEC regime" $\Delta=-10\mu$.  The points are
raised a distance above the $k$-$\omega$ plane proportional to the
height of the delta-function peaks.  These large values of $\Delta$
are chosen to emphasize the important qualitative features.}
\label{sf}
\end{figure}


As we show below, the RF spectra observed by Schunck et al.
\cite{Schunck} are consistent with a traditional Fermi liquid
scenario,
where the low energy excitations can be placed in one-to-one correspondence with those of an ideal Fermi gas.
 The higher energy excitations of this 
system are however quite unusual.  We find an anomaly in the
down-spin spectral density at energies/momenta around the up-spin
Fermi energy/momentum.  This spectral feature, characterized by a
sudden bending of the spectrum, effectively separates the single
particle spectrum into two branches, yielding the two peaks seen in
experiments.
%

Before discussing a theory of the spin imbalanced normal state it is
helpful to review the physics of RF spectroscopy in mean-field
theory of the superfluid state. For simplicity we will neglect the
interaction between the atoms in the $|{\rm ex}\rangle$ state and
the other two spin states.  This approximation should not be
quantitatively correct for the recent experiments, where there are
several Feshbach resonances in close proximity \cite{lithium}.
Despite the importance of understanding these final state effects
\cite{gupta}, one does not expect that they play any qualitative
role in the observations. Under these assumptions, particles in the
excited internal state have a free dispersion $\epsilon_{\rm
ex}(k)=\epsilon_{\rm ex}(0)+k^2/2m$, and Fermi's golden rule  tells
us that the number of atoms transfered by an RF spectroscopy
experiment is proportional to \cite{Zoller,Griffin,kinnunen}
\begin{equation}\label{spectrum}
I(\nu)=\int \frac{d^3k}{(2\pi)^3} A_\downarrow(k,k^2/2m+\mu_\downarrow -\nu) n_\downarrow(k^2/2m+\mu_\downarrow +\nu),
\end{equation}
 where $\nu$ is the detuning of the radio frequency field from the free-space splitting between the $|{\rm ex}\rangle$ state and the $\downarrow$ state,
$n(\epsilon)=(e^{\beta \epsilon}+1)^{-1}$ is the Fermi function,
$\beta$ is the inverse temperature, and the spectral density
$A_\downarrow(k,\omega)$ represents how many down-spin single
particle
 states exist at a given momentum and energy (measured from the down-spin chemical potential).
   Figure~\ref{sf} shows the mean-field result $A(k,\omega)=v_k^2 \delta(\omega-E_k)$,
   where the coherence factor is $v_k^2=(\epsilon_k-E_k)/(2 E_k)$, the noninteracting spectrum
   is $\epsilon_k=k^2/(2m)-\mu$ and the quasiparticle spectrum is $E_k^2=\epsilon_k^2+\Delta^2$,
   with the superfluid gap given by $\Delta$.  The spectral density is qualitatively different in the BCS regime
   ($\mu>0$) and the BEC one ($\mu<0$).  Qualitatively, the unitary gas is similar to the BCS
   limit as it has $\mu>0$.  It is important to note that aside from near the places where the
   two branches become close, most of the spectral weight lies near the free particle spectrum $\omega = \epsilon_k$.

These mean-field spectra
can be understood by noting that within the mean-field picture
there are two
%
ways to add an down-spin particle with momentum $k$.  One can
directly add it (costing energy $\epsilon_{k\downarrow}$) or one can
remove an up-spin particle of momentum $q-k$ and simultaneously add
a pair with momentum $q$. In this latter case the change in energy
is the difference between the energy of the pair measured from the
pair chemical potential $E_b+q^2/2m_b-\mu_b$, and the energy of the
up-down particle which was removed. In the superfluid state there is
a condensate of $q=0$ pairs, requiring $E_b-\mu_b=0$. One therefore
finds that the energy to add the spin-down particle in this manner
costs energy $-\epsilon_{k\uparrow}$. An avoided crossing between
these two modes gives the BCS spectrum. In the BEC limit there is no
avoided crossing, and one just has two parabolic bands.

%

\begin{figure}[!tb]
\includegraphics[height=2.3in,width=3.4in,angle=0]{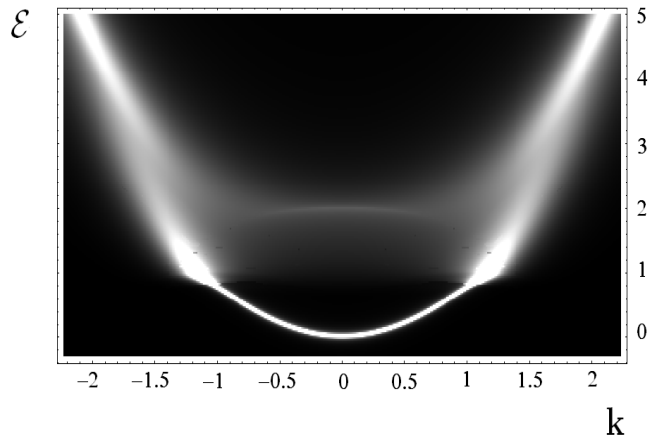}
\caption{Spectral density of spin down quasiparticles
$A(\epsilon,k)$ at zero temperature. The $\epsilon$ axis is
normalized by $\mu_{\uparrow}$, the $k$ axis by $\sqrt{2 m
\mu_{\uparrow}}$, $m$ being the bare mass of atoms. The value of the
chemical potential ratio in this plot is
$\mu_{\downarrow}/\mu_{\uparrow} = -0.6$, the scattering length is
taken to be infinite.}\label{sd}
\end{figure}

In contrast to the mean field treatment of the superfluid, when one
includes fluctuations in the normal state (either above $T_c$ or in
the normal state of a spin imbalanced gas), there are many possible
channels for adding a single down-spin particle. In principle, one
should consider all possible ways of exciting the many body state
while adding the particle. Below we will take into account processes
where a single spin-up particle hole pair is created. Including all
such processes leads to anomalous features in the spectral density
of spin-down particles. The spectral density is a nonanalytic
function of energy and momenta at the values where the phase space
for the creation of particle-hole pairs hits the boundary of the
particle-hole continuum. This feature, which is a consequence of the
sharp Fermi Surface, corresponds to Kohn anomalies which are
observed in the spectra of phonons \cite{kohn}. At momenta of the
order of the Fermi momentum of spin up particles, the spectrum of
spin down particles bends and becomes damped, creating a dip in the
density of states. This effectively translates into a separation of
the spectrum into a low momentum (shifted down from the free
spectrum) and a high momentum branch (shifted up from the free
spectrum), which gives to the density of states in the normal state
(Fig. 2) a structure similar to that in the superfluid state (Fig.
1).

Geometrically, the RF spectroscopy intensity $I(\nu)$ can be found
by overlaying the parabola $k^2/2m - \nu+\mu_\downarrow$ on the
down-spin spectral density graph and quantifying their overlap. Note
that since momentum is conserved under absorption of an RF photon,
this experiment is very different from a point-contact tunneling
experiment in solid state physics. Rather it is equivalent to a
solid state photoemission experiment (if one does not
momentum-resolve the final state) or a tunneling experiment in which
momentum is conserved (such as occurs in tunneling between parallel
wires \cite{yacoby}).

In the superfluid state at finite temperature one sees a bimodal RF
spectrum as each of the two branches of the single particle spectrum
contributes to the RF lineshape at different frequencies.  Since the
upper branch asymptotically approaches that of a free gas, one of
the two spectral lines begins at $\nu=0$. The separation $\delta$
between the two lines is given by the difference between the $k =0$
energy of the free parabola and the $k=0$ energy of the lower
branch: $\delta=\sqrt{\mu_\downarrow^2+\Delta^2}-\mu_\downarrow$.
(It is particularly important to realize that this splitting is not
simply the gap $\Delta$ \cite{Baym}.)  At high temperature the
contribution from the upper branch dominates, while at low
temperature the contribution from the lower branch dominates.  On
the BCS side of resonance, where pairing mostly occurs near the
Fermi surface, both peaks are dominated by unpaired states. As is
clear from the spectral density shown in Figure~\ref{sf}(a), the
presence of pairs only plays a role very near to the gap. On the
other hand, on the BEC side of resonance, the entire lower branch is
due to pairs. Clearly, the existence of a shifted low temperature RF
peak is not sufficient to conclude that the system is superfluid, a
counterexample being a gas of fermions with interaction treated in
the Hartree Fock approximation.

\begin{figure}[!tb]
\includegraphics[height=2.4in,width=3.4in,angle=0]{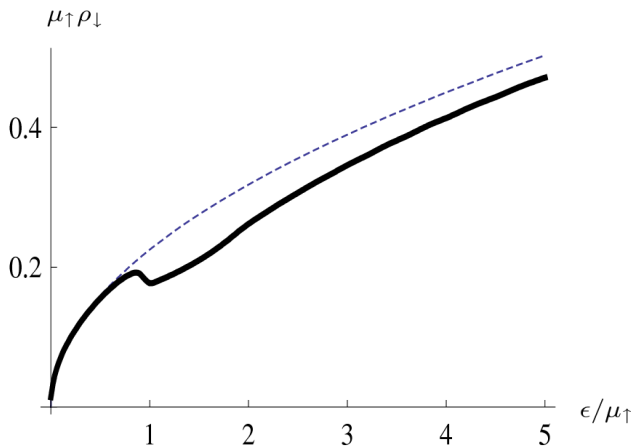}
\caption{ Single particle density of states
$\rho_\downarrow(\epsilon)$, normalized by $\mu_{\uparrow}$, plotted
in the absence (dashed) and in the presence (solid) of interaction
as a function of $\epsilon/\mu_{\uparrow}$. The density of states is
reduced relative to the noninteracting value near
$\epsilon\sim\mu_\uparrow$.} \label{dos}
\end{figure}

\begin{figure}[!tb]
\includegraphics[height=4.8in,width=3.2in,angle=0]{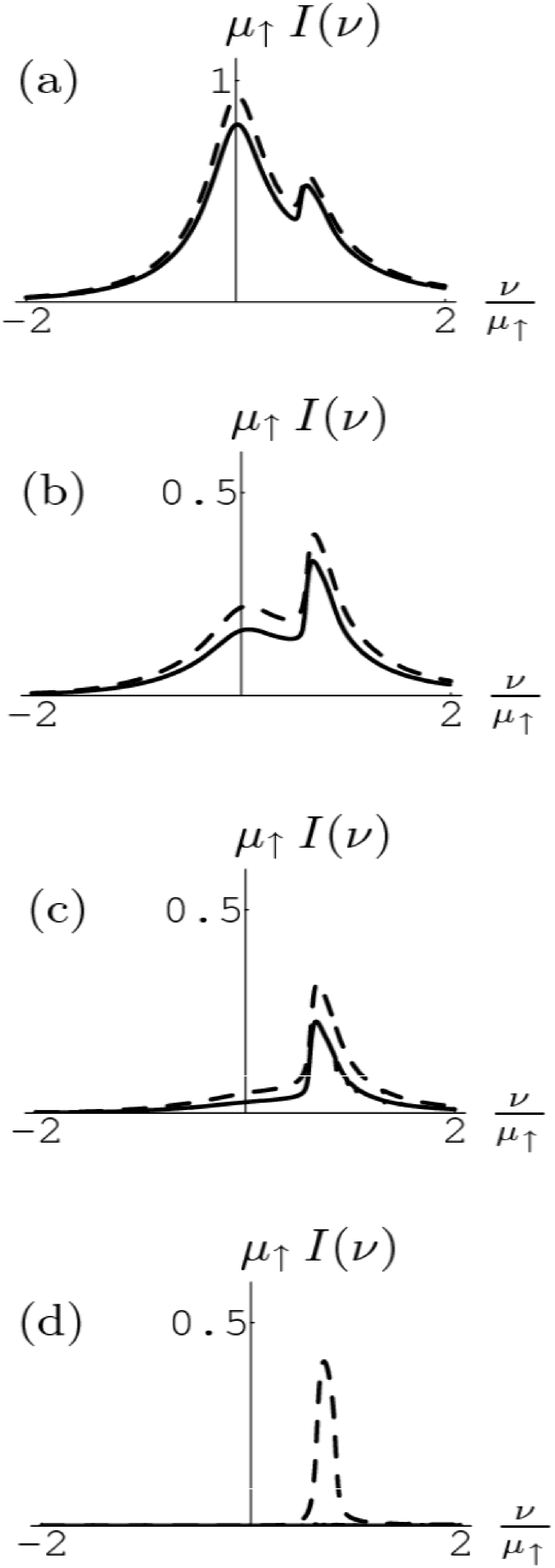}
\caption{ Calculated RF spectrum in the normal state at unitarity:
(a) $T=4\mu_\uparrow$, (b) $T=2\mu_\uparrow$, (c) $T=\mu_\uparrow$
(d) $T=0$.  Solid lines correspond to
$\mu_\downarrow/\mu_\uparrow=-0.60671$,  resulting in vanishing
minority species density at $T=0$, Dashed lines correspond to
$\mu_\downarrow/\mu_\uparrow=0.1$. } \label{peaks}
\end{figure}

To extend this picture to the spin imbalanced gas we must calculate
the spectral density $A(k,\omega)$. Since there exists to our
knowledge no controlled method for calculating the properties of a
Fermi gas at unitarity, we work with the simplest approximation
which captures the basic physics -- namely the many-body T-matrix
approximation which generalizes Noziere and Schmidt-Rinks theory of
the BCS-BEC crossover to the spin imbalanced gas.  Variants of this
approximation have been used by several groups
\cite{levin,Strinati}. We use the one described by Combescot et al
\cite{Combescot1,Combescot2}, who have shown that this approximation
gives remarkably good agreement with Monte-Carlo calculations in the
limit of vanishing downspin density. Specifically we take
\begin{eqnarray}
A_\downarrow(\omega, k) &=& \mathrm{Im} \  \left(\omega -\epsilon_{k}^{\downarrow} - \Sigma_{\downarrow}(\omega, k)\right)^{-1}\\
\Sigma_\downarrow(\omega, k) &=&\int dz\, \Gamma_\downarrow(z,k)/(2\pi (\omega-z))\\
\Gamma_\downarrow(\omega,k)&=&\int_{0<\epsilon_k<\omega} d^3q/(2\pi)^3\Lambda(\omega+\epsilon_{q}^{\uparrow},k+q)\\
\Lambda(\omega,k)&=&2{\rm Im}T(\omega,k)\\
T(\omega,k)&=& (4\pi \hbar^2/m)/(a^{-1} +\Theta(\omega,k))\\
\Theta(\omega,k)&=& \int \frac{dz}{2\pi(\omega-z)}\int\frac{d^3q}{(2\pi)^3}\\\nonumber&&
\left[\frac{1-f^\uparrow_{k/2+q}-f^\downarrow_{k/2-q}}
{\omega-\epsilon_{k/2+q}^\uparrow-\epsilon^\downarrow_{k/2-q}}
-\frac{m}{k^2}\right]
\end{eqnarray}
where $\epsilon^{\sigma}_k=k^2/2m-\mu_\sigma$, and $f^{\sigma}_k=\theta(-\epsilon^{\sigma}_k)$.
Note that spectra used to calculate the self-energy are free spectra.
 Self-consistency should not qualitatively change the spectral density,
 and may actually make the theory less accurate \cite{kadanoffmartin}.

The resulting spectral density is shown in Fig.~\ref{sd}, where the
separation between a small momentum branch, underdamped and
parabolic, and a branch at higher momenta, shifted upward, is
clearly recognizable.  This two branch structure is similar to the
mean-field spectral densities in Fig.~\ref{sf}, except for the fact
that the branches are joined by an overdamped continuum.  In the
limit of vanishing $N_\downarrow$, (occuring at
$\mu_\downarrow=-0.6067$) we find that the lower branch is described
by $E(k)=k^2/2m^*$, with an effective mass $m^*=1.16$.
%
At sufficiently large momenta, the damping of quasiparticles is
vanishingly small. This is because the fundamental interaction is
short-ranged, and unitarity of the S-matrix requires that scattering
becomes weak at large momenta. Figure ~\ref{dos} shows the
integrated density of states
\be \rho(\epsilon) = \frac{1}{2 \pi^2} \int_0^{\infty} d p  \ p^2
A(p,\epsilon) \ee
 where a dip is clearly visible at energies close
to the Fermi level. Here, we mention that a similar dip is found in the
spectral density of cuprate superconductors \cite{levin}.


In Figure 4, we use Eq. 1 to calculate the RF-spectroscopy
lineshape. Note that since we are using the zero-temperature
spectral density the finite temperature line-shapes are at most
qualitative. The general structure, however is generic. At the
lowest temperatures only the bottom branch of the spectrum is
occupied, and one sees only a single peak, shifted from $\delta=0$
by an amount proportional to $\mu_\uparrow$. In the limit of
vanishing downspin density this shift is directly equal to the
downspin chemical potential $\mu_\downarrow=\Sigma(k=0,\omega=0)$,
and provides a model independent way to determine this quantity.
As temperature rises the upper branch becomes occupied, resulting in
a second peak at lower frequencies.  The weight in the second peak
grows with the temperature, eventually dwarfing the low temperature
peak.

We believe that in a finite temperature calculation of $A(k,\omega)$
one would find that the separation between the two branches would
become smaller as temperature increased.
This would manifest itself in the low temperature
peak slowly moving towards lower energy,
merging with the high temperature peak.
The splitting should vanish at a characteristic temperature $T^*\sim \mu_\uparrow/k_B$.

We are grateful to the Institute Henri Poincar\'{e} and to the
Workshop on Quantum Gases for hospitality. F.F. acknowledges
discussions with C. Schunck and G. Shlyapnikov, and financial
support from the Marie Curie Contract MEST-CT-2005-0197555.  E.J.M
acknowledge support from NSF grant PHY-0456261 and the Alfred P.
Sloan Foundation.

\end{document}